\documentclass[aps,prl,reprint,superscriptaddress,longbibliography]{revtex4-2}
\usepackage{graphicx}
\usepackage{color}
\usepackage[tight]{units}
\usepackage{epstopdf}
\usepackage[normalem]{ulem}
\DeclareGraphicsExtensions{.pdf,.jpg,.png,.eps}
\usepackage{amsfonts}
\usepackage{amsmath}
\usepackage{amssymb}
\usepackage{color}
\usepackage{color}
\usepackage{lineno}
\usepackage{epstopdf}
\usepackage{adjustbox}

\usepackage[T1]{fontenc}
\usepackage{hyperref}
\newcommand{\xf}{(Ni$_{0.2}$Mg$_{0.2}$Co$_{0.2}$Cu$_{0.2}$Zn$_{0.2}$)(MnFe)O$_{4}$}

\begin{document}
\title{Ultra-Soft Ferrimagnetism in a High-Entropy Spinel Oxide Driven by Site-Selective Cation Disorder}
\author{Neha Sharma}
\affiliation{Department of Physics and Material Science, Thapar Institute of Engineering and Technology, Patiala 147004, India}
\author{AmritPal}
\affiliation{UGC-DAE Consortium for Scientific Research, Khandwa Road, Indore 452001, Madhya Pradesh, India}
\author{Nikita Sharma}
\affiliation{Department of Physics and Material Science, Thapar Institute of Engineering and Technology, Patiala 147004, India}
\author{Mathieu Duttine}
\affiliation{Univ. Bordeaux, CNRS, Bordeaux INP, ICMCB, UMR 5026, Pessac, F-33600, France}
\author{Denis Pelloquin}
\affiliation{Laboratory Crismat, UMR6508 CNRS, Normandie University, ENSICAEN, UNICAEN, 6 bd Maréchal Juin, 14050 Caen cedex 4, France}
\author{S. D. Kaushik}
\affiliation{UGC-DAE Consortium for Scientific Research, Mumbai Centre, 246-C CFB, BARC Campus, Trombay, Mumbai-400085, India}
\author{Sanjoy Mahatha}
\affiliation{UGC-DAE Consortium for Scientific Research, Khandwa Road, Indore 452001, Madhya Pradesh, India}
\author{Olivier Toulemonde}
\affiliation{Univ. Bordeaux, CNRS, Bordeaux INP, ICMCB, UMR 5026, Pessac, F-33600, France}
\author{Sourav Marik}
\email[]{soumarik@thapar.edu}
\affiliation{Department of Physics and Material Science, Thapar Institute of Engineering and Technology, Patiala 147004, India}

\begin{abstract}
High-entropy materials are complex, multifunctional materials that have reshaped the design of advanced functional materials. Their chemically diverse compositions enable access to a broader compositional space than conventional solid solutions, while simultaneously posing significant challenges for fundamental structure–property understanding.
In this study, we introduce a new high-entropy spinel oxide with an exceptionally low coercivity of 1.8 Oe at room temperature, among the lowest reported for bulk spinel oxides, and a high electrical resistivity (1560 ohm-cm). Neutron powder diffraction (NPD) and magnetic measurements reveal long-range collinear ferrimagnetic ordering (k = 0,0,0) with a transition temperature at 420 K. This rare combination of ultra-soft magnetic behavior, robust ferrimagnetic ordering well above room temperature, and high resistivity highlights its strong potential as an advanced soft-magnetic oxide for low-loss, high-frequency applications. Furthermore, X-ray absorption spectroscopy (XAS), Mössbauer spectroscopy, and NPD analyses were combined to determine the cation distribution and site selectivity across the tetrahedral and octahedral sites of the complex structure.

\end{abstract}
\maketitle
\section{Introduction}
The emerging high-entropy strategy has revolutionized the design of functional materials by incorporating five or more elements in equiatomic or near-equiatomic ratios within the cation sublattices \cite{1, 2, 3, 4, 5}. This approach significantly enhances the chemical complexity and disorder, leading to extraordinary properties distinct from those of conventional materials. High-entropy oxides (HEOs) have developed as a promising class of materials as a result of their exceptional structural and functional versatility \cite{6}. This multi-component configuration leads to a high configurational entropy, which plays a crucial role in stabilizing the solid solution. For an oxide of the general formula $A_{\alpha}B_{\beta}O_{\gamma}$, the configurational entropy $S_{\text{config}}$ is expressed as \cite{7}
\begin{align}
S_{\text{config}} = -R \Bigg[ \alpha \sum_{i=1}^{N} a_i \ln a_i \notag 
 + \beta \sum_{j=1}^{M} b_j \ln b_j \notag 
& + \gamma \sum_{k=1}^{P} c_k \ln c_k 
\Bigg]
\label{eq:Sconfig}
\end{align}
Here, $a_i$, $b_j$, and $c_k$ are the mole fractions of elements occupying the A-site, B-site, and O-site respectively, and $\alpha$, $\beta$, $\gamma$ are the corresponding stoichiometric coefficients in the oxide formula. This entropy-driven stabilization facilitates the formation of complex oxides with tunable magnetic \cite{8}, catalytic \cite{9} and electronic functionalities \cite{10}.
High entropy oxides (HEOs) have been synthesized in various crystal structures, including perovskite \cite{11}, fluorite \cite {12}, spinel \cite{13}, and pyrochlores \cite{14}, typically forming single-phase solid solutions when configurational entropy dominates over enthalpy, leading to entropy stabilization. The first experimentally realized entropy-stabilized oxide, (Mg$_{0.2}$Co$_{0.2}$Ni$_{0.2}$Cu$_{0.2}$Zn$_{0.2}$)O, \cite{7, 15} was introduced by Rost et al. in 2015 and crystallized in a rock-salt structure. The stability of such materials is governed by a thermodynamic balance wherein the increase in configurational entropy ($\Delta$S) compensates for the positive enthalpy change ($\Delta H$), resulting in a negative Gibbs free energy ($\Delta G$ = $\Delta H$- T$\Delta$S). This delicate interplay between high entropy and positive enthalpy represents a defining feature of entropy-stabilized oxides. The ability to tailor material properties by leveraging compositional and configurational complexity makes HEOs an exciting platform for developing novel functionalities, including electrical \cite{16, 17, 18}, thermal \cite{19}, catalytic \cite{9, 20}, optical \cite{21}, and magnetic properties \cite{22, 23, 24}.

In this study, we introduced a new ultra-soft ferrimagnetic high entropy spinel oxide with composition \xf. Spinel materials are particularly attractive because of their broad range of technological applications. Their dual-cation sublattice structure and intrinsic structural flexibility have placed spinels as promising candidates for tuning a wide range of physico-chemical properties through complex compositional and entropy engineering \cite{25, 26, 27}. 
 In recent years, a surge in the research interest has emerged in high-entropy spinel oxides, as reflected by several studies such as the oxidative polymerization of water contaminants oxygen evolution reactions (OER) \cite{13, 28}, solid oxide fuel cells (SOFCs) \cite{29}, tunable and entropy-protected magnetism \cite{30, 31}  complex and spin glass-like magnetism \cite{32, 33, 34} enhanced exchange bias \cite{35}, and robust ferrimagnetic behaviour \cite{36}.\\
 
 The diversity of magnetic responses in these systems is mainly governed by both local and extended spin arrangements, which define the variety of metal–oxygen–metal exchange interactions present in the complex structure \cite{37, 38, 39}. Therefore, key factors such as coordination geometry, valence and spin states, cationic composition, and inherent structural disorder collectively influence the magnetic tunability. A recent study by Wang et al. \cite{40} on Mg–Fe–Co–Ni–Cu–Mn–Zn spinel ferrites highlights this complexity, showing that cation inversion significantly affects bulk magnetization and spin correlations, further emphasizing the role of entropy-mediated local-structural modifications in tailoring magnetic functionalities.
However, the precise cation distribution in spinel oxides is a long-standing challenge, particularly in high-entropy spinel oxides, where multiple principal cations with several oxidation and spin states coexist. A detailed understanding of site-specific cation occupancy is essential, as it directly influences both the structural framework and magnetic interactions in these complex oxides. Herein, we have explored an extremely soft ferrimagnetic spinel high entropy material \xf. Employing a multi-pronged experimental approach involving X-ray and neutron powder diffraction (NPD), M\"ossbauer spectroscopy, and X-ray absorption spectroscopy (XAS), we probe the oxidation states and site selectivity of individual cations. 

\begin{figure*}
\includegraphics[width=18cm,height=14cm,scale=1.5]{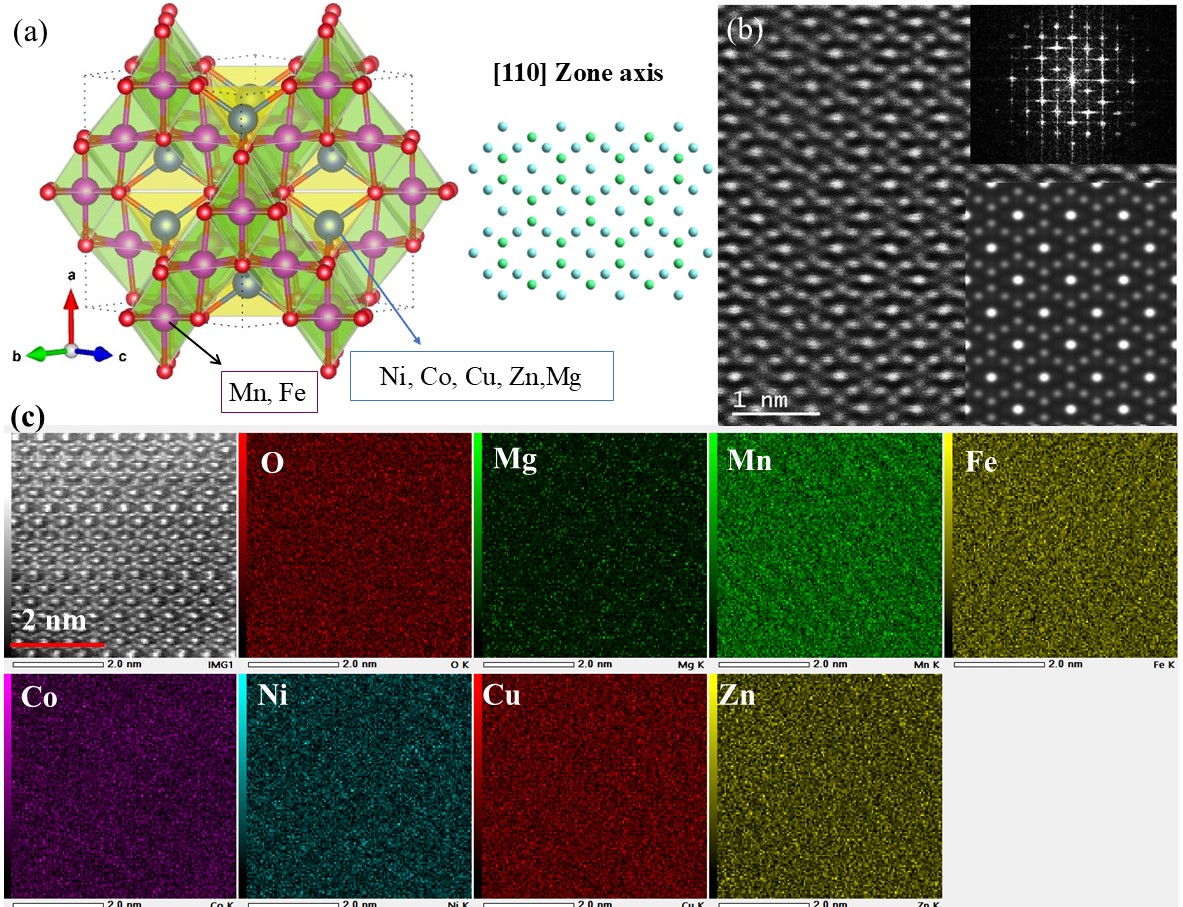}
\centering
\caption{(a) Crystal structure of the cubic spinel phase with [110] oriented cation projection: blue circles are related to B atomic rows and green circles are related to A atomic rows considering the ideal AB$_2$O$_4$ spinel structure, 
(b) Experimental [110] HAADF image and corresponding fourier form transform (FFT) pattern. A simulated HAADF image is inserted in the right part of the image. (c) Nano-scale EDS elemental mapping showing homogeneous distribution of elements for \xf collected at room temperature.}
  \label{plot 2}
\end{figure*}

\begin{figure*}[ht]
\includegraphics[width=1.00\linewidth]{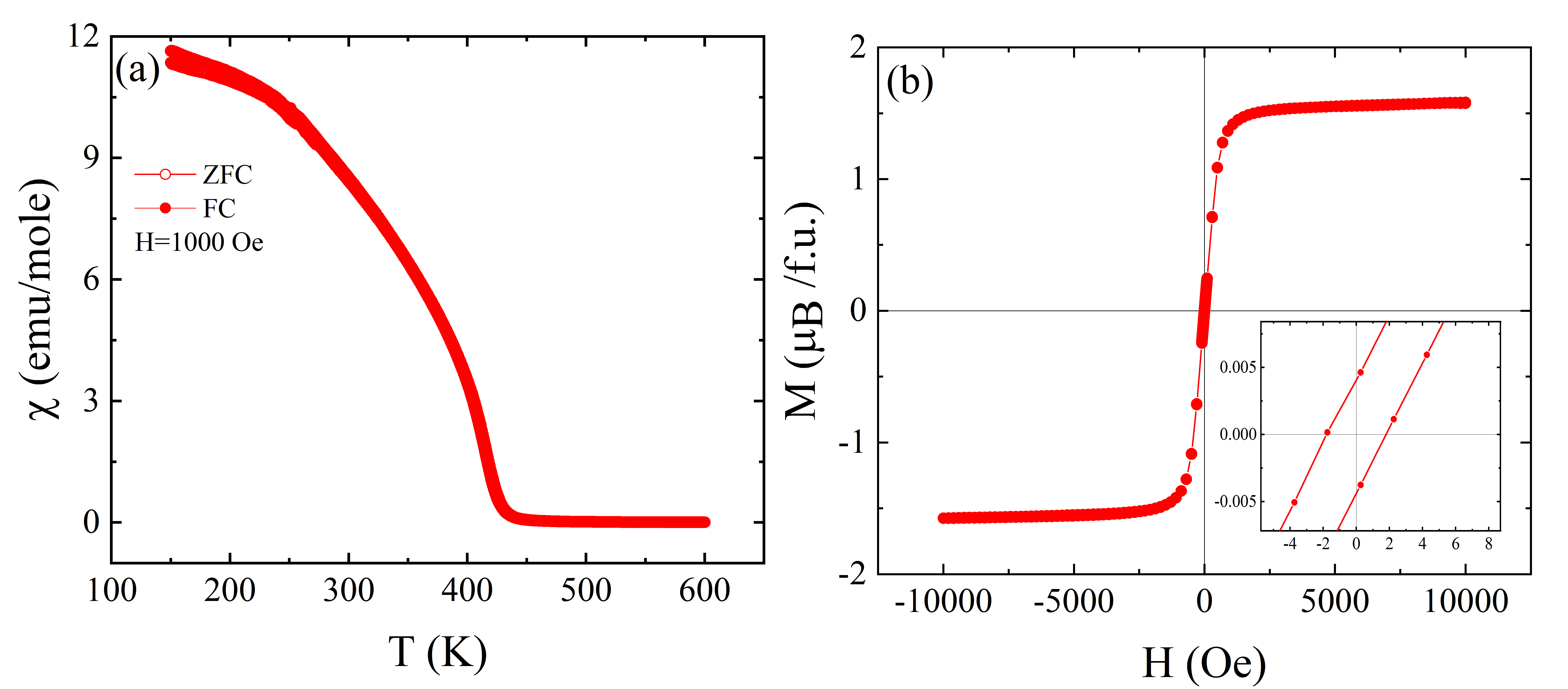}
\caption{(a) Temperature-dependent magnetic susceptibility (\textit{$\chi$} vs. \textit{T}) curves of \xf measured under an applied field of 1000 Oe, showing zero-field-cooled (ZFC) and field-cooled (FC) behavior, and (b) magnetic field dependence of magnetization (M-H loop) measured at 300 K. Inset shows the enlarged part of the M-H loop, highlighting the extremely low coercivity.}
\label{plot 3}
\end{figure*}

\section{Results and Discussion}
The synthesis conditions and other experimental details are provided in the supporting information (SI) file. The initial phase identification and structural analysis of the synthesized material were performed using room-temperature powder X-ray diffraction (XRD, Fig. S1). The XRD results confirm the cubic crystal structure \textit{Fd-3m} with a = b = c = 8.4039 (4) \AA. A detailed crystal structure analysis is carried out using a neutron powder diffraction pattern collected at 10 K, taking into account the site selectivity of various cations as determined from M\"ossbauer and XAS measurements. The results are discussed in detail in the latter part of the manuscript. To gain deeper insight into the chemical composition, morphology, and microscale chemical homogeneity, we have performed scanning electron microscopy (SEM) and nano-scale (using Scanning Transmission
Electron Microscopy) energy-dispersive X-ray spectroscopy (EDS) measurements at room temperature. The SEM image (Fig. S2(c) in the SI) reveals a compact arrangement of polyhedral grains with clearly defined boundaries, indicating effective grain growth during high-temperature sintering \cite{41}. The corresponding size distribution shows an average grain size of about 4.8 $\mu$m, indicating uniform particle formation throughout the sample (Fig.  S2 in the SI File). Importantly, atomic-level EDS elemental mapping (Figure \ref{plot 2} (c)) confirms an even distribution of all cations at the nanometer scale, highlighting the chemical homogeneity of the synthesized material.  Furthermore, point-wise EDS measurements across different regions (Table S1, SI file) reflect the target composition. Figure \ref{plot 2} (b) shows the high-angle annular dark-field Scanning Transmission Electron Microscopy (HAADF-STEM) image acquired along the [110] zone axis, revealing well-defined atomic columns.  The Fourier-transformed (FFT) pattern highlights cubic symmetry. A simulated HAADF image is inserted in the right part of the image.
The observed lattice contrast and periodic arrangements are consistent with the spinel crystal structure.

To further probe the local structural environment and its response to different A- and B-site cations (AB$_2$O$_4$ structure, Fig. 1 (a)), Raman spectroscopy is performed. Raman spectra provide valuable information on phonon modes, which are highly responsive to variations in cation type, mass, and bonding configuration within the spinel lattice. In our Raman spectra (Fig. S1b in the SI file), four distinct bands are identified at 176, 307, 489, and 632 cm\textsuperscript{-1},  closely resembling the phonon features of conventional spinels like  NiFe$_{2}$O$_{4}$ \cite{42} and other high entropy cubic spinel oxides \cite{43}. Group theory predicts that spinel structures with the \textit{Fd-3m} space group should exhibit five-order Raman-active modes as follows:
\begin{equation}
\quad A_{1g} (R) + E_g (R) + 3F_{2g} (R)
\label{eq1}
 \end{equation}
 All Raman-active modes in spinels are associated with the motion of oxygen atoms (Equ. \ref{eq1}) \cite{42}. Specifically, in cubic spinels, the A$_{1g}$ mode (600–720 cm$^{-1}$) corresponds to vibrations along the bond between an oxygen atom and the tetrahedral M$^{2+}$ cation. Similarly, the E$_{1g}$ mode (250–360 cm$^{-1}$) arises from the symmetric bending of the oxygen anion relative to the octahedral M$^{3+}$ cation. In addition, the F$_{2g}$(1) mode (160–220 cm$^{-1}$) is attributed to the translational motion of M$^{2+}$ with oxygen, whereas the higher-frequency F$_{2g}$(2) and F$_{2g}$(3) modes (440–590 cm$^{-1}$) are linked to the asymmetric stretching and bending of oxygen respectively. The observed reduction in the number of Raman bands, along with asymmetric peak shapes and a distinct shoulder near 700 cm$^{-1}$ {A$_{g}$ (1) suggests deviations from the ideal spinel symmetry. The occurrence of the soldier near 700 cm$^{-1}$ indicates the presence of multiple cations at both sites in the spinel structure. The peak broadening indicates a substantial degree of disorder within the lattice, aligning with the statistical occupancy of various cations at crystallographic sites, a characteristic of configurational entropy in HEOs \cite{24,41,44}.

DC magnetization measurements were conducted under an applied field of 1000 Oe in both zero-field-cooled (ZFC) and field-cooled (FC) modes, as shown in Figure ~\ref{plot 3}. The magnetization exhibits a sharp increase below 420 K, indicative of a long-range ferrimagnetic ordering transition. The sharpness of the magnetic transition suggests a robust collective magnetic state, which may be stabilized by the high configurational entropy, as reported in other high-entropy spinel systems \cite{31}. The negligible bifurcation between the ZFC and FC curves suggests minimal magnetic frustration, despite high disorder, indicating a well-established ferrimagnetic phase.

The isothermal magnetization curve (M-H) is recorded at 300 K, with a step size of 2 Oe in the low-field region.  Figure ~\ref{plot 3}(b) shows a well-defined hysteresis loop with a high saturation magnetization (M$_s$) (\(M_s = 8834\) emu/mole, 38.4 emu/gm) and extreme-low coercivity of 1.8 Oe (H$_C$). In comparison, other reported low entropy and high entropy bulk spinel oxides exhibit typical coercivities above 10 Oe \cite{45, 46,47}. For instance, Mg$_{0.5}$Ni$_{0.5}$FeMnO$_4$ shows a coercivity value of 34 Oe, Mg$_{0.33}$Ni$_{0.33}$Cu$_{0.33}$FeMnO$_4$ show H$_C$ = 28 Oe (Fig. S5 in the SI). Similarly, high entropy Mg$_{0.2}$Ni$_{0.2}$Co$_{0.2}$Cu$_{0.2}$Zn$_{0.2}$Fe$_2$O$_4$, Mg$_{0.2}$Ni$_{0.2}$Co$_{0.2}$Fe$_{0.2}$Cu$_{0.2}$Fe$_2$O$_4$, Mg$_{0.2}$Ni$_{0.2}$Co$_{0.2}$Mn$_{0.2}$Cu$_{0.2}$Fe$_2$O$_4$, Mn$_{0.2}$Ni$_{0.2}$Co$_{0.2}$Fe$_{0.2}$Cu$_{0.2}$Fe$_2$O$_4$, and [CoFeCrNiMn]$_3$O$_4$ show coercivity above 15 Oe \cite{25}. Therefore, the present material exhibits a significantly reduced coercivity. To the best of our knowledge, among oxides, the current composition exhibits the lowest reported coercivities for a bulk sample at room temperature, indicating ultra-soft ferrimagnetic behavior. At the same time, the observed room temperature saturation magnetization value is comparable with the other spinel materials \cite{45,46}. The moderate saturation magnetization at room temperature could be due to strong superexchange interactions mediated through the oxygen lattice among the B-site cations \cite{48, 49}. The exceptionally low coercivity observed here is not a generic consequence of either spinel ferrimagnetism or configurational entropy. Conventional spinel oxides (including the inverse spinels) generally contain a limited number of cation species occupying each crystallographic sublattice, whereas the high-entropy spinel accommodates multiple principal cations (five or more) within the same crystallographic environment, generating an entropy-stabilized disordered landscape. As mentioned earlier, previously reported conventional and high-entropy spinel oxides exhibit substantially larger coercive fields. In the present material, the experimentally determined site-selective cation arrangements (discussed in the later part of the manuscript) appear to be associated with the observed ultra-soft magnetic behavior. To further investigate this relationship, we systematically modified the cation-disorder landscape by substituting Mn with Cr (magnetic) and Al (non-magnetic). These substitutions resulted in progressively larger coercive fields (Fig. S9, S10 in the SI file), increasing from 1.8 Oe in the parent composition to approximately 2.5 Oe and 3.9 Oe for the partially Cr- and Al-substituted compositions, respectively, and to 6 Oe for the fully Cr-substituted sample. In comparison, the corresponding Fe-rich composition exhibits H$_C$ $\approx$ 35 Oe (Fig. S11 in the SI file). The systematic increase in coercivity upon modifying the cation arrangement suggests that the ultra-soft ferrimagnetic state is highly sensitive to the specific site-selective disorder realized in the parent composition. In addition, resistivity measurements reveal a high room-temperature resistivity of  1564 ohm-cm (see SI, Fig. S3). The coexistence of high saturation magnetization, extremely low coercivity, and high electrical resistivity is particularly significant from an application perspective. Such a combination is highly desirable for soft magnetic insulating core materials used in alternating-current (AC) magnetic devices and memory devices. The ultra-low coercivity minimizes hysteresis loss per cycle, while the high resistivity suppresses eddy current losses under time-varying magnetic fields. Consequently, the present material is a strong candidate for high-frequency transformer cores, power inductors, EMI suppression components, and miniaturized power electronics operating in the kHz–MHz regime.  

The extremely low coercivity suggests a substantial reduction in magnetocrystalline anisotropy and a minimized density of domain-wall pinning centers. The reduction in coercivity can be correlated with cation distribution, inversion degree, and intrinsic configurational disorder \cite{50}. To further investigate the magnetic microstructure associated with the ultra-soft ferrimagnetic behavior, room-temperature magnetic force microscopy (MFM, Fig. S6) measurements were carried out. The MFM phase image exhibits relatively smooth and spatially extended magnetic contrast without signatures of strong magnetic pinning sites. The continuous domain morphology indicates a reduced effective pinning landscape and supports facile domain-wall motion, consistent with the exceptionally low room-temperature coercivity (1.8 Oe). Also, in high-entropy spinels, the random occupation of crystallographic sites by multiple magnetic cations could lead to a statistical averaging of local anisotropy contributions, and that could lower the effective magnetocrystalline anisotropy. Further experiments, such as by torque magnetometry or ferromagnetic resonance, are required to understand the exact role of anisotropy in this sample. However, a detailed understanding of cation distribution and site occupancy is essential to elucidate the disorder-induced enhancement of magnetic softness in this complex spinel oxide. In the next sections, we have used several element specific tools to understand the cationic distributions and site selectivity in this complex spinel oxides.

At room temperature (295 K), the $^{57}$Fe M\"ossbauer spectrum ($H_{\text{ext}} = 0$) exhibits a not fully-ordered magnetic component (Figure S4), confirming that the magnetic transition occurs close to room temperature, in agreement with the magnetization data.
At low temperature (5 K), the spectrum (Figure 3) can be clearly reconstructed with two magnetic sextets associated with two distinct Fe environments, both corresponding to high-spin (HS) Fe$^{3+}$ ions. Refined hyperfine parameters (isomer shift ($\delta$), quadrupole line shift ($2\epsilon$) and hyperfine magnetic field $B_{\text{hf}}$) are summarized in the SI file (Table S2). The light blue sub-spectrum, with a relative area of 68\%, represents Fe$^{3+}$ ions occupying octahedral sites, characterized by $B_{\text{hf}} = 50.9(8)~\text{T}$. The deep blue sub-spectrum, with a relative area of 31\% and $B_{\text{hf}} = 50.2(5)~\text{T}$, corresponds to Fe$^{3+}$ ions at tetrahedral sites.

\begin{figure}[h!]
    \centering
    \includegraphics[width=0.5\textwidth]{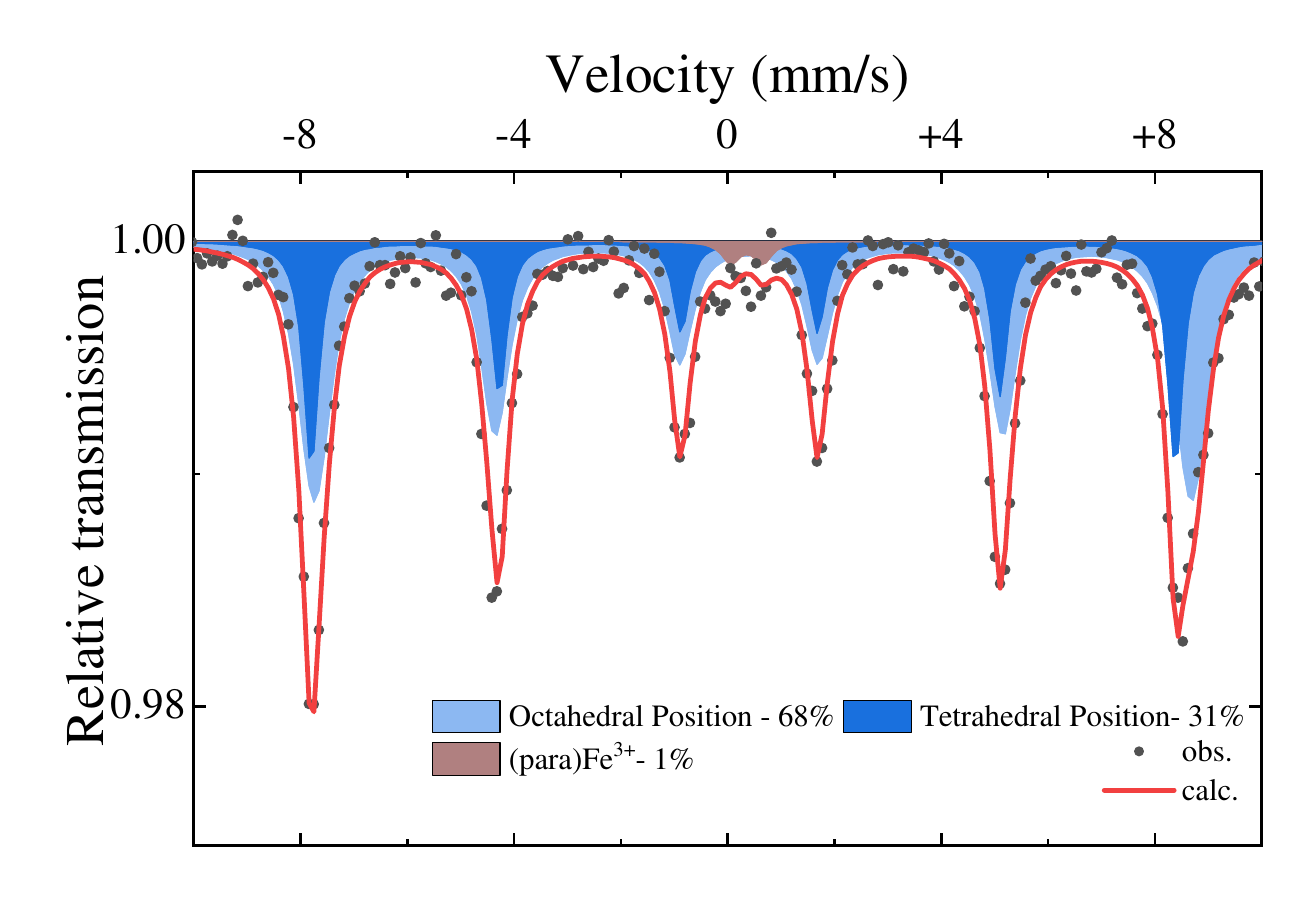}
    \caption{Low temperature (5 K) $^{57}$Fe M\"ossbauer spectrum of the high-entropy oxide sample. The spectrum shows Fe$^{3+}$ ions distributed over octahedral (68\%) and tetrahedral (31\%) sites, with a minor paramagnetic Fe$^{3+}$ component (1\%). Experimental data (black dots) and fitted curve (red line) confirm the magnetic ordering and cation site distribution typical of high-entropy oxides.}
    \label{plot 4}
\end{figure}
\begin{figure*}[ht]
    \centering
\includegraphics[width=2.0\columnwidth]{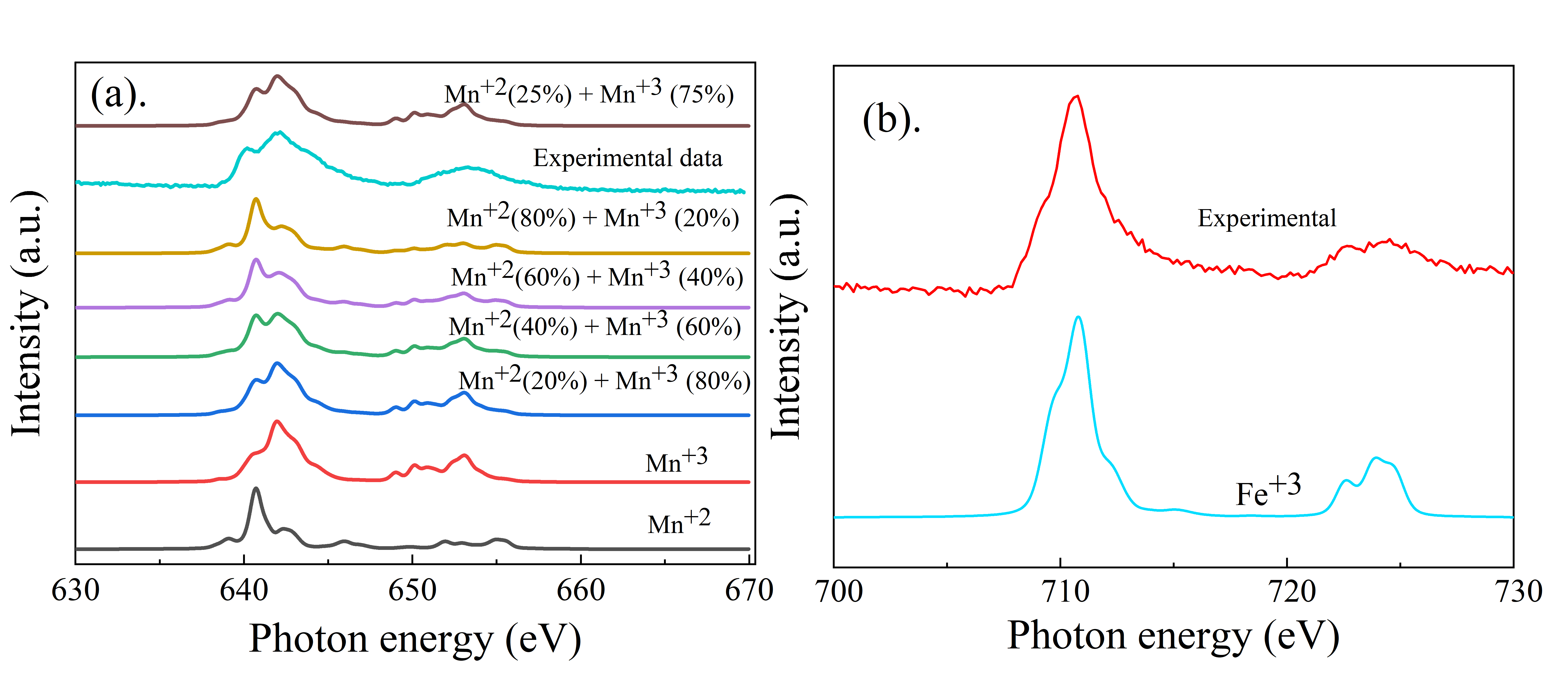}
\caption{Room temperature experimental and simulated (a) Mn L edge and (b) Fe$_{2,3}$ edge XAS spectra for the sample.}
\label{plot 5}
\end{figure*}

\begin{table}[h]
\small
\centering
\caption{Atomic and structural parameters obtained from Rietveld refinement of 10 K neutron powder diffraction (NPD) data for (Ni$_{0.2}$Mg$_{0.2}$Co$_{0.2}$Cu$_{0.2}$Zn$_{0.2}$)(MnFe)O$_4$.}
\label{tab:NPD10K}
\begin{tabular}{lll}
\hline
\textbf{Parameter} & \textbf{T = 10 K} & \textbf{} \\
\hline
\textbf{Space Group} & \textit{Fd$\bar{3}$m} & Cubic spinel \\
$a = b = c$ (\AA) & 8.4084 (1) & Lattice parameter \\
\hline
\multicolumn{3}{l}{\textbf{A-site Cations (0.125, 0.125, 0.125)}} \\
Fe 1 Occupancy & 0.22 (2)  & B$_\text{iso}$ = 0.81 (3) \\
Mn 1 Occupancy & 0.17 (2) & B$_\text{iso}$ = 0.81 (3) \\
Mg Occupancy & 0.2 & B$_\text{iso}$ = 0.81 (3) \\
Cu Occupancy & 0.2 & B$_\text{iso}$ = 0.81 (3) \\
Zn Occupancy & 0.2 & B$_\text{iso}$ = 0.81 (3) \\
\hline
\multicolumn{3}{l}{\textbf{B-site Cations (0.5, 0.5, 0.5)}} \\
Fe 2 Occupancy & 0.37 (2) & B$_\text{iso}$ = 0.23 (2) \\
Co   Occupancy & 0.1 & B$_\text{iso}$ = 0.23 (2) \\
Mn 2 Occupancy & 0.42 (2) & B$_\text{iso}$ = 0.23 (2) \\
Ni Occupancy & 0.1 & B$_\text{iso}$ = 0.23 (2) \\
\hline
\multicolumn{3}{l}{\textbf{Oxygen (x, x, x)}} \\
x & 0.26026 (5) & Oxygen positional parameter \\
Occupancy & 1 & Fully occupied \\
B$_\text{iso}$ & 0.93 (1) & -- \\
\hline
B--B distance (\AA) & 2.97286(3) & Octahedral edge \\
A--B distance (\AA) & 3.48598(3) & Tetra--octahedral \\
$\chi^2$ & 1.92 &  \\
$R_p$ & 3.41 &  \\
$R_{wp}$ & 3.24 &  \\
$R_{mag}$ & 3.42 &  \\
M(T$_d$) ($\mu_B$) & 2.849 (8) & A-site magnetic moment \\
M(Oct.) ($\mu_B$) & 1.906 (8) & B-site magnetic moment \\
\hline
\end{tabular}
\end{table}

To refine the 10 K NPD pattern and guide the site selectivity of cations, we used XAS and M\"ossbauer spectroscopy results, along with the concept of octahedral site preference energy (OSPE). Figure~\ref{plot 5} displays the Fe and Mn L-edge X-ray Absorption Spectra (XAS, collected at room temperature), accompanied by simulated reference spectra used to evaluate the oxidation states and coordination environments of these transition metal cations. XAS is a highly sensitive, element-specific technique that probes the unoccupied density of states, offering direct insight into the local chemical environment, including oxidation state, coordination geometry, and electronic structure. The Mn L-edge spectrum [Figure~\ref{plot 5}(a)] was analyzed by comparing the experimental data with simulated spectra for Mn$^{3+}$ in octahedral coordination and Mn$^{2+}$ in tetrahedral sites. These simulations were generated using the \texttt{CTM4XAS} package, which integrates atomic multiplet theory, crystal field theory, and charge transfer effects to model electronic transitions \cite{51}. Through iterative fitting of mixed-valence states, the best match to the experimental spectrum was achieved using a composition of approximately 75\% Mn$^{3+}$ (octahedral) and 25\% Mn$^{2+}$ (tetrahedral), indicating a predominantly trivalent manganese character with a minor divalent contribution. The simulation parameters were optimized with values of 10$Dq$ = 1.2 (1.3), $F_{dd}$ = 1.5 (0.8), $F_{pd}$ = 1.5 (0.8), and $G_{pd}$ = 1.5 (0.8), with Lorentzian and Gaussian broadenings of 0.2 and 0.4, respectively, for Mn$^{3+}$ (Mn$^{2+}$). Similarly, the Fe L-edge spectrum [Figure~\ref{plot 5}(b)] exhibits spectral features characteristic of Fe$^{3+}$. The asymmetric peak shape further suggests multiplet splitting, consistent with trivalent iron, and is in good agreement with the M\"ossbauer studies. However, due to spectral overlap and intrinsic resolution limitations at the L-edge, precise quantification of Fe site occupancy (in octahedral and tetrahedral sites) remains challenging from the present XAS measurements. The Co L-edge spectra (Figure S5, SI) indicate the elements are in the 2+ oxidation state. However, the present data do not allow an accurate determination of their site selectivity. Also, previous studies on high-entropy spinel oxides have highlighted that octahedral site preference energies (OSPEs) play a role in governing cation distribution \cite{26}. Based on OSPE considerations, Ni is expected to preferentially occupy octahedral sites.

 \begin{figure}
\includegraphics[width=1.0\columnwidth]{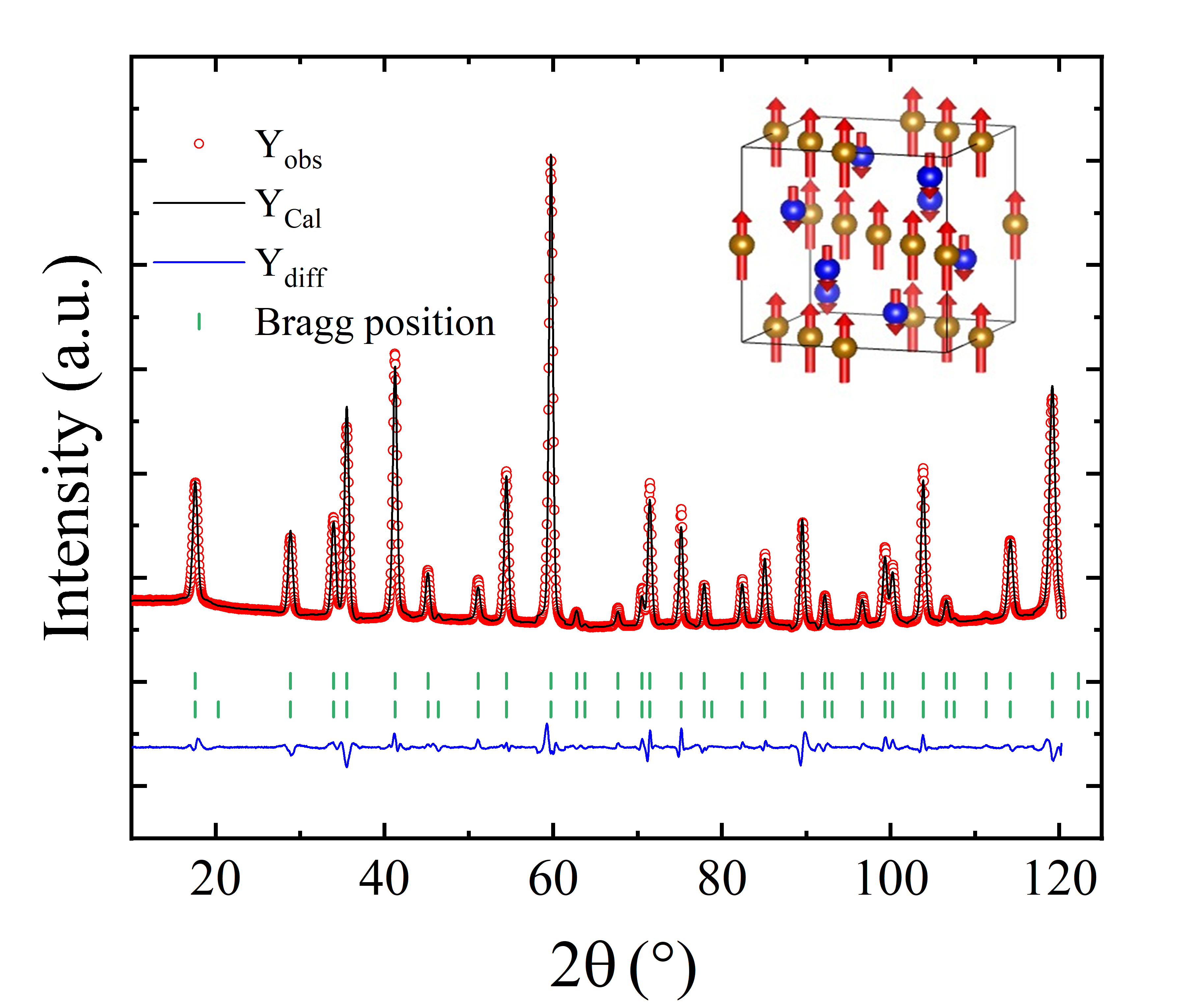}
\centering
\caption{Rietveld refinement of the neutron powder diffraction pattern collected at 10 K for \xf. The lower tick marks indicate the positions of magnetic reflections with the propagation vector k = (0, 0, 0). The inset illustrates the magnetic structure, with green spheres representing A-site cations and blue spheres representing B-site cations.}
  \label{plot 6}
\end{figure}

Figure~\ref{plot 6} presents the Rietveld refinement of neutron powder diffraction (NPD) data collected at 10 K for the synthesized high-entropy spinel oxide. The refinement confirms that the sample crystallizes in a single-phase cubic spinel structure with the space group \textit{Fd-3m} and a lattice parameter of $a = 8.4084 (1)$~\AA. The magnetic structure obtained from the refinement is shown in the inset of Figure \ref{plot 6}.
Refinement of the oxygen sublattice showed no deviation from full occupancy, indicating a stoichiometric oxide structure. Due to the distinct neutron scattering lengths of the constituent elements (e.g., Ni: 1.03\,fm, Fe: 0.945\,fm, Mn: -0.373\,fm, Co: 0.249\,fm O: -0.58030), neutron diffraction proves effective in resolving site-specific occupancy and detecting oxygen vacancies. 
To guide site assignments, we employed XAS and M\"ossbauer spectroscopy results, along with the concept of {octahedral site preference energy} (OSPE) ~\cite{26}. Based on OSPE values, Ni prefers to occupy octahedral B-sites (16d), while Mg, Cu, and Zn were refined on the tetrahedral A-sites (8a). Fe and Mn were allowed to occupy both sites (Table 1). The best fit to the diffraction data was obtained with the chemical formula: [Mg$_{0.2}$Cu$_{0.2}$Zn$_{0.2}$Mn$_{0.17}$Fe$_{0.22}$]
 [Ni$_{0.1}$Mn$_{0.42}$Fe$_{0.37}$Co$_{0.1}$]$_2$O$_{4}$, consistent with our M\"ossbauer and XAS measurements.

 Despite the severe disorder and chemical complexity, the 10\,K NPD data clearly exhibit long-range magnetic ordering. All magnetic reflections are indexed with a propagation vector \textbf{k} = (0, 0, 0), suggesting a collinear ferrimagnetic structure, analogous to those reported in CoFe$_2$O$_4$ \cite{52}. The absence of the (200) magnetic Bragg reflection at $2\theta = 20.35^\circ$ rules out non-collinear spin arrangements such as Yafet-Kittel-type canting. In contrast, the prominent (400) reflection at $2\theta = 41.21^\circ$ supports the collinear nature of the spin alignment. The refined magnetic moments at the tetrahedral (M$_\mathrm{Td}$) and octahedral (M$_\mathrm{Oct}$) sites are listed in Table 1. A notable reduction in the observed magnetic moments, relative to theoretical values, is attributed to spin dilution and local canting effects induced by the presence of non-magnetic (Mg$^{2+}$, Zn$^{2+}$) and weakly magnetic (Cu$^{2+}$) cations.
 
\section{Conclusion}
In this work, we systematically investigated the site-specific cation distribution and occupancy in a complex high-entropy spinel oxide using a comprehensive multi-technique experimental approach. The combined use of X-ray diffraction, X-ray absorption spectroscopy, neutron powder diffraction, scanning electron microscopy, scanning transmission electron microscopy, energy-dispersive X-ray spectroscopy and M\"ossbauer spectroscopy enabled a detailed and correlated understanding of the oxidation states, coordination geometries, and spatial distribution of transition-metal cations across the tetrahedral (A) and octahedral (B) sites of the spinel lattice. Microscopic and spectroscopic analyses (SEM, STEM, nanoscale EDS) confirmed homogeneous elemental distribution and single-phase microstructure with long-range structural ordering. XAS provides insight into valence states and local coordination environments, while NPD provides quantitative information on cation site occupancies. M\"ossbauer spectroscopy further probed the local electronic and magnetic environments of Fe species. The integration of these complementary techniques establishes the chemical composition as [Mg$_{0.2}$Cu$_{0.2}$Zn$_{0.2}$Mn$_{0.17}$Fe$_{0.22}$]
 [Ni$_{0.1}$Mn$_{0.42}$Fe$_{0.37}$Co$_{0.1}$]$_2$O$_{4}$ and confirms that the incorporation of multiple cations within the spinel framework leads to intrinsic configurational disorder characteristic of high-entropy systems. Low-temperature NPD, M\"ossbauer spectroscopy, and magnetic measurements collectively reveal robust long-range collinear ferrimagnetic ordering (k = 0,0,0) with a magnetic transition temperature well above room temperature. Remarkably, the material exhibits an exceptionally low coercivity of 1.8 Oe at 300 K, among the lowest reported for bulk spinel oxides, alongside moderate saturation magnetization and high electrical resistivity (1560 ohm-cm). The ultra-soft ferrimagnetic behavior is likely associated with reduced effective pinning arising from chemically disordered cation environments. The site-selective cation disorder associated with high-entropy cationic arrangements plays a crucial role in facilitating a facile domain-wall motion and producing ultra-soft ferrimagnetic behavior. Overall, the entropy-stabilized ultra-soft ferrimagnetic state demonstrated here highlights the effectiveness of cation disorder engineering in tailoring magnetic and functional performance. 

\section*{Acknowledgements} \par S. M. acknowledges financial support from the ANRF, Government of India, for the Start-up Research Grant (SRG/2021/001993) and Advanced Research Grant (ANRF/ARG/2025/004870/PS). S. M. and O. T. acknowledge the financial support from the SPARC, Ministry of Education, Govt. of India (Project no. P4139). N. S. acknowledges financial support from the Council of Scientific and Industrial Research (CSIR), Government of India, under CSIR HRDG Ref. No. 09/0677(23323)/2025-EMR-I. D.P. acknowledges the French national Transmission Electron Microscopy and Atomic Probe (METSA) network for the TEM facilities and the PAMEC platform for the technical support.

\section*{Conflicts of interest}
There are no conflicts to declare.

\section*{Data availability}
All the data can be made available upon a reasonable request from the corresponding author.

\end{document}